\documentclass[conference]{IEEEtran}
\IEEEoverridecommandlockouts
\usepackage{cite}
\usepackage{amsmath,amssymb,amsfonts}
\usepackage{algorithmic}
\usepackage{graphicx}
\usepackage{textcomp}

\usepackage{tablefootnote}

\usepackage{multirow}
\usepackage{multicol}
\usepackage{soul,xcolor}
\usepackage[normalem]{ulem}

\graphicspath{{figures/}} 

\begin{document}

\title{Deep Reinforcement Learning for Task Offloading in UAV-Aided Smart Farm Networks}
\author{\IEEEauthorblockN{Anne Catherine Nguyen$\dag$, Turgay Pamuklu$\dag$, \IEEEmembership{Member, IEEE}, Aisha Syed$\ddag$, \\ W. Sean Kennedy$\ddag$, Melike Erol-Kantarci$\dag$, \IEEEmembership{Senior Member, IEEE}}

\IEEEauthorblockA{$\dag$\textit{School of Electrical Engineering and Computer Science,}
\textit{University of Ottawa}, Ottawa, Canada}

\IEEEauthorblockA{$\ddag$\textit{Nokia Bell Labs}\\
Emails:\{anguy087, turgay.pamuklu, melike.erolkantarci\}@uottawa.ca, 
\{aisha.syed, william.kennedy\}@nokia-bell-labs.com}
}

\maketitle
\makeatletter
\def\ps@IEEEtitlepagestyle{%
  \def\@oddfoot{\mycopyrightnotice}%
  \def\@oddhead{\hbox{}\@IEEEheaderstyle\leftmark\hfil\thepage}\relax
  \def\@evenhead{\@IEEEheaderstyle\thepage\hfil\leftmark\hbox{}}\relax
  \def\@evenfoot{}%
}
\def\mycopyrightnotice{%
  \begin{minipage}{\textwidth}
  \centering \scriptsize
Accepted Paper. IEEE policy provides that authors are free to follow funder public access mandates to post accepted articles in repositories. When posting in a repository, the IEEE embargo period is 24 months. However, IEEE recognizes that posting requirements and embargo periods vary by funder. IEEE authors may comply with requirements to deposit their accepted manuscripts in a repository per funder requirements where the embargo is less than 24 months.
  \end{minipage}
}
\makeatother

\begin{abstract}
The fifth and sixth generations of wireless communication networks are enabling tools such as internet of things devices, unmanned aerial vehicles (UAVs), and artificial intelligence, to improve the agricultural landscape using a network of devices to automatically monitor farmlands. Surveying a large area requires performing a lot of image classification tasks within a specific period of time in order to prevent damage to the farm in case of an incident, such as fire or flood. UAVs have limited energy and computing power, and may not be able to perform all of the intense image classification tasks locally and within an appropriate amount of time. Hence, it is assumed that the UAVs are able to partially offload their workload to nearby multi-access edge computing devices. The UAVs need a decision-making algorithm that will decide where the tasks will be performed, while also considering the time constraints and energy level of the other UAVs in the network. In this paper, we introduce a Deep Q-Learning (DQL) approach to solve this multi-objective problem. The proposed method is compared with Q-Learning and three heuristic baselines, and the simulation results show that our proposed DQL-based method achieves comparable results when it comes to the UAVs' remaining battery levels and percentage of deadline violations. In addition, our method is able to reach convergence 13 times faster than Q-Learning.   
\end{abstract}

\section{Introduction}
Smart agriculture has emerged as an efficient way to manage large farmlands. It uses a collection of Internet of things (IoT) devices as sensors to monitor the land and the environmental changes, and utilizes artificial intelligence (AI) to make precise farming decisions. The sensors are deployed in various locations across the farm, they collect information from their surroundings, and then forward the information toward a server. The AI agent then uses the information from the server to make informed decisions in order to take actions to protect or improve the efficiency of the farm. To provide the promised bandwidth and ultra-low latency specifications, 5G and 6G networks will depend on unmanned aerial vehicles (UAVs) as a source to provide untethered and ubiquitous mobile connectivity and computing services through the air and reach remote rural areas.  


An example of an AI-based technique that can aid in the decision-making process for farms, is image classification. It can be used to detect pests, monitor crop growth, or detect fires. In \cite{Aldabbagh2020}, Aldabbagh et al. proposed a Deep Learning algorithm to identify the growth stages of chili plants. They used a dataset of chili plant images in various growth stages to train a Deep Neural Network (DNN). Similarly in \cite{Lina2020}, Yu et al. used image classification to identify pest infestations on crops from images that were captured by UAVs. With the introduction of smart agriculture, farmers can use IoT devices with sensors, such as cameras, to monitor their crops and use machine learning-based image classification algorithms, to monitor pests, fires, and the growth stages of crop.”

Accurate image classification is a computationally intensive task because it may require running a DNN. Even though the IoT devices are necessary for providing regular updates on the farm, they are limited in computing capacity. As a result, they do not have the required capacity to perform the image classification tasks in a timely manner and will need to offload it to a nearby device in order to complete the task. This is crucial because the farm environment can change quickly and failing to respond to the changes within a certain time frame can cause detrimental damages. For example, a fire can cause minimal damage if it is detected and put out early. However, a farmer can lose a substantial amount of crops if the fire has not been detected for several minutes.  

A network consisting of UAVs with mounted base stations and Multi-Access Edge Computing (MEC) devices has been proposed to aid in farm monitoring tasks in the literature. Zhao et al. \cite{Zhao2020} used such a network in order to monitor a farm. The UAVs and MEC are both equipped with a processing unit that is capable of performing image classification tasks. The base stations on the UAVs allow the IoT devices to connect with more processing units that are able to perform the intensive image classification tasks. The tasks can be forwarded to a UAV or to a MEC device. The objective of \cite{Zhao2020} is to optimize the delay of the data processing and transmission by finding the optimal resource allocation and UAV trajectory.  

With a three-dimensional range of motion, UAVs offer a wide range of services. They can provide line of sight connectivity between the IoT devices, and other UAVs or MEC devices. They can also provide computing capabilities. However, UAVs are limited by their battery capacities, If the UAV is performing too many image classification tasks, their hover time will be greatly reduced. \cite{Zhang2020} and \cite{Li2019} both proposed that in a 5G and beyond network, MEC devices can alleviate the UAV's workload while also improving the task's latency. 

In this paper, we are using Deep Q-Learning (DQL) to improve the performance of the algorithms found in \cite{Nguyen2021}. Similarly, we consider a smart farm with IoT cameras, UAVs, and a MEC server. Our objective is to find a task offloading solution that is both energy efficient and that respects the deadlines of the tasks. In a nutshell, the objectives of this study are: 1) to find a task scheduling policy that will: elongate the UAVs' hover times, and minimize the number of tasks that do not meet their deadlines, and 2) to decrease the convergence time to the optimal solution. We compare our DQL-based solution with several benchmarks including Q-learning and heuristic algorithms. We show that DQL outperforms the benchmark solutions.

The rest of this paper is organized as follows. Section II introduces the related works. Section III describes the problem and the system model. Section IV describes the proposed method and baseline algorithms. Section V describes the simulation results and analysis. Finally, Section VI is the conclusion and discussion of future works.

\section{Related Work}
Using reinforcement learning (RL) to manage wireless network resources in order to optimize performance is widely studied across many different applications. In \cite{Elsayed2019b}, Elsayed et al. surveyed the challenges and opportunities of AI in 5G and 6G networks. They introduced multiple applications for RL in wireless networks such as energy management and radio resource allocation. They also identified that using AI for energy efficiency would be essential in 6G. Furthermore, Khoramnejad et al. \cite{Khoramnejad2021} proposed a deep RL approach to solve a joint optimization problem consisting of maximizing computation and minimizing energy consumption for a 5G and beyond network through offloading. Their network also makes use of MEC servers as a processing unit to assist their network in computing-intensive tasks. Likewise, Akbari et al. \cite{Akbari2021} introduced a deep RL algorithm in an industrial IoT environment. The algorithm aimed to find an optimal placement and scheduling policy for virtual network functions in order to minimize end-to-end delay and cost.

There have been several studies using UAVs in a smart farm. Lottes et al. \cite{Lottes2017} detailed how UAVs can be used to capture aerial images, and image classification is used to identify the crops and weeds in the field. In the survey paper \cite{Islam2021}, Islam et al. introduced the idea of using UAVs to spray pesticide, and discussed the trade-off between latency and battery usage.

The usage of both UAVs and MEC devices is beneficial for applications in 5G and beyond networks. In \cite{Zeng2019}, Zeng et al. provided a survey on the benefits and challenges of integrating UAVs in a 5G and beyond network. Fonseca et al. discussed the challenges for the network operators when UAVs are integrated into a network \cite{Fonseca2021}. \cite{Zhang2020}, \cite{Zhou2020}, and \cite{Li2019} have provided extensive surveys on the use of UAVs with MEC for different applications such as space-air-ground networks, and emergency search and rescue missions. In addition, Hu et al. \cite{Hu2021} discussed using UAVs to provide connectivity for 6G internet of vehicles applications. 

The existing approaches to optimize energy consumption and latency for UAVs are not limited to smart farm scenarios. For instance, Yang et al. \cite{Yang2019} aimed to reduce power consumption by optimizing the following parameters ``user association, power control, computation capacity allocation and location planning''. In \cite{Zhou2021}, Zhou et al. considered a network that consists of satellites, UAVs, terrestrial base stations, and IoT devices. They used deep RL as a task scheduling solution that minimizes processing delay while considering the UAVs' energy constraints. Alternatively, Ghdiri et al. \cite{Ghdiri2020} used clustering and trajectory planning in order to optimize energy efficiency and tasks' delay time. Additionally, Yao et al. \cite{Yao2020} used a game theory solution for the task offloading problem in a UAV swarm scenario. Although we are exploring a similar problem we focus on jointly solving the energy and task latency optimization problem through DQL.

\section{System Model}

\par Our network consists of a set of UAVs, $j\in\mathcal{J}$. They can communicate with IoT devices $z \in\mathcal{Z}$, other UAVs, and a set of MEC servers  $l\in\mathcal{L}$. Every UAV has a battery with a maximum capacity of $\Upsilon^{B}_{j}$. Both UAVs and MEC devices have processing capabilities, $j'\in \mathcal{J}^{+}$, where they can process the IoT devices' tasks.

\begin{figure}
    \centering
    \includegraphics[width=0.48\textwidth, height=5cm]{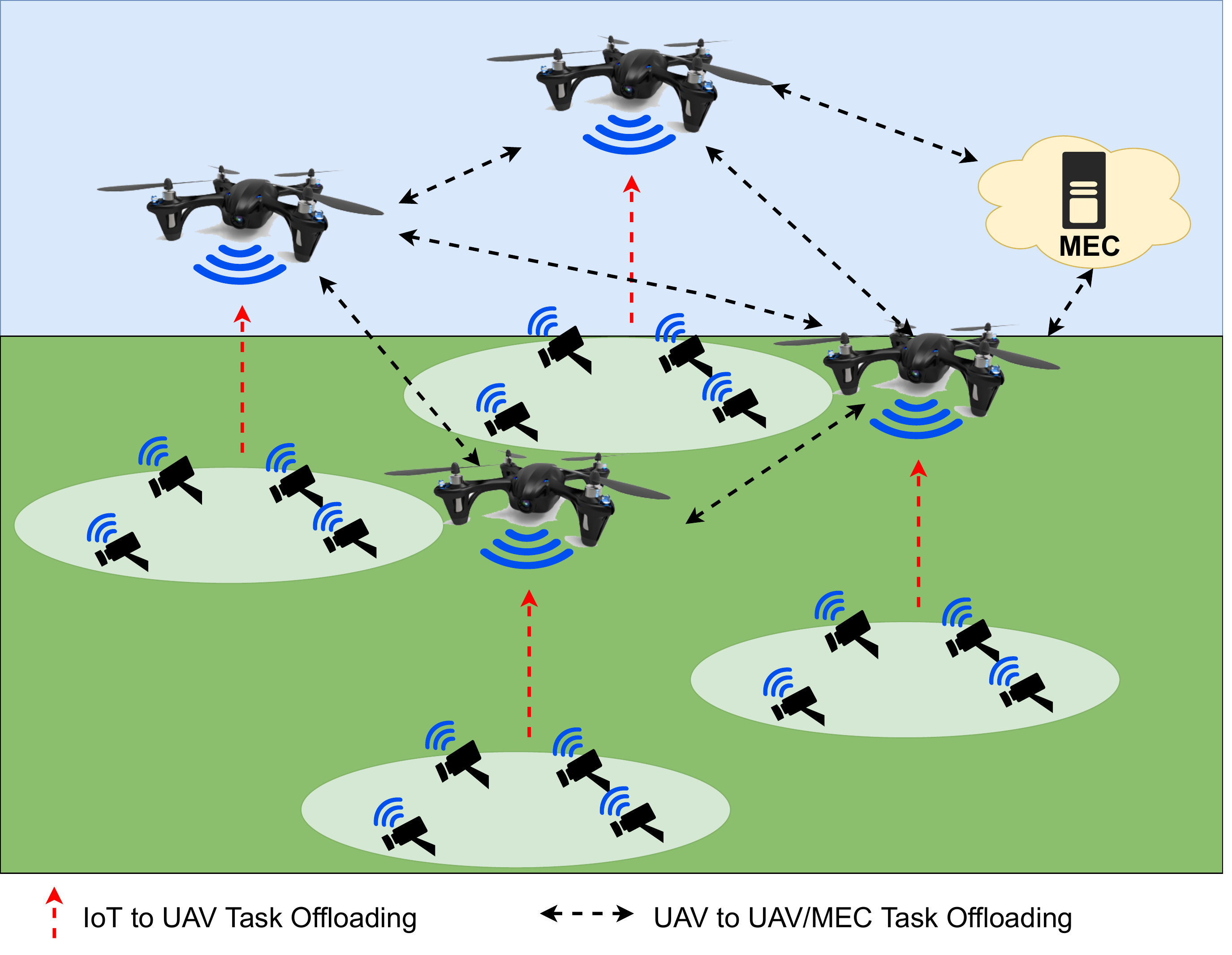}
    \caption{\label{fig:sysmodel} Overview of smart farm network}
\end{figure}

As displayed in Fig. \ref{fig:sysmodel}, at time $t\in \mathcal{T}$, the IoT devices can offload $K$ types of tasks to a UAV ($\alpha^{B}_{jt}$). Each task type has a predefined deadline $\alpha^{D}_{jt}$, and the amount of time it takes for the processing unit to execute such a task $\alpha^{P}_{jt}$. The goal of this paper is to find a scheduling algorithm for each UAV to assign each task to a processing unit in a way such that the tasks can be completed before their deadline, and the UAVs' hover time will be maximized \cite{Nguyen2021}. These two objectives are combined to form our multi-objective maximization problem,

\textbf{Maximize: }
\begin{flalign}
\label{eq:obj}
&W * \min_{ j' \in J}\Upsilon^{R}_{j'} - \frac{1-W}{\Theta} \sum\limits_{\substack{j\in\mathcal{J}\\ t\in\mathcal{T}}}v_{jt},
\end{flalign}

where $W$ refers to the importance of the maximizing hover time objective, $\Upsilon^{R}_{j'}$ refers to a UAV's remaining battery level, $v_{jt}$ refers to the number of deadline violations that have occurred, and $\Theta$ refers to the scaling factor used to normalize $v$. The first goal is to maximize the lowest remaining battery level, in order to extend the UAV network's hover time. 

The UAV's remaining battery level $\Upsilon^{R}_{j'}$ can be calculated as follows,
\begin{flalign}
\label{eq:encalc}
\Upsilon^{R}_{j'} =& \Upsilon^{B}_{j'} - (\Upsilon^{H}_{j'} + \Upsilon^{A}_{j'} + \Upsilon^{I}_{j'}) * \mathcal{T} \notag \\ 
& - \sum\limits_{\substack{j\in\mathcal{J}\\ t\in\mathcal{T} \\ t'\in\mathcal{T}}} (\Upsilon^{C}_{j'} - \Upsilon^{I}_{j'}) * p_{jtj't'},
\end{flalign}

where $\Upsilon^{B}_{j'}$ is the battery capacity, $\Upsilon^{H}_{j'}$ is the amount of energy required for the UAV to hover, $\Upsilon^{A}_{j'}$ is the amount of energy the antenna requires to transmit a signal, $\Upsilon^{I}_{j'}$ is the amount of energy consumed by the processing unit in idle mode, $\mathcal{T}$ is the simulation time, and $\Upsilon^{C}_{j'}$ is the amount of energy consumed by the processing unit while the processing unit is performing a task. Finally, $p_{jtj't'}$ is a binary decision variable that equals one if processing unit $j'$ processes a task.

The processing unit delay $\Delta_{jt}$ is the total number of times that a task must remain in the processing unit's queue plus the task's processing delay $\alpha^{P}_{jt}$. Processing unit delay is given by,
\begin{flalign}
\label{eq:CPUdelay}
\Delta_{jt} = \sum\limits_{\substack{j\in\mathcal{J}^{+} \\ t\in\mathcal{T}}} \left[p^{+}_{jtj't'} * (t') - t + \alpha^{P}_{jt} \right],
\end{flalign}

where $p^{+}_{jtj't'}$ is a binary decision variable that equals one if it is the time interval that processing unit $j'$ has started processing the task, $t'$ is the time interval that the task has started on processing unit $j'$, and $t$ is the time interval in which the task has arrived at processing unit $j$. 

A deadline violation, $v_{jt}$, occurs at $t$ when the sum of the IoT to UAV transmission delay, $\Delta^{z}_{jt}$, processing unit delay, $\Delta_{jt}$, and the transmission delay between processing units, $\Delta^{jt}_{j'}$ exceeds the task's deadline, $\alpha^{D}_{jt}$. It can be formulated as,
\begin{flalign}
\label{eq:DV}
v_{jt} = 
\begin{cases}
        1 \text{ when } \Delta^{z}_{jt} + \sum_{j'\in\mathcal{J}^{+}}x_{jtj'} * \Delta^{jt}_{j'} + \Delta_{jt} > \alpha^{D}_{jt} \\
        0 \text{ when } \Delta^{z}_{jt} +  \sum_{j'\in\mathcal{J}^{+}}x_{jtj'} * \Delta^{jt}_{j'} + \Delta_{jt} \leq \alpha^{D}_{jt} 
\end{cases}
\end{flalign}

where $ x_{jtj'} $ is used to determine if the task was done at processing unit $j'$. It is set to 1 when the task is going to be performed at processing unit $j'$, otherwise, it will be set to 0.

In order to avoid the ping-pong effect, a task can only be offloaded one time,
\begin{equation}
\label{eq:offloadlimit}
\sum\limits_{j'\in\mathcal{J}^{+}} x_{jtj'}  \leq 1 ,\quad \forall j\in\mathcal{J}, \forall t\in\mathcal{T}.
\end{equation}	

\section{Proposed Method}

\subsection{Deep Q-Learning}
\begin{figure}
    \centering
    \includegraphics[width=0.5\textwidth]{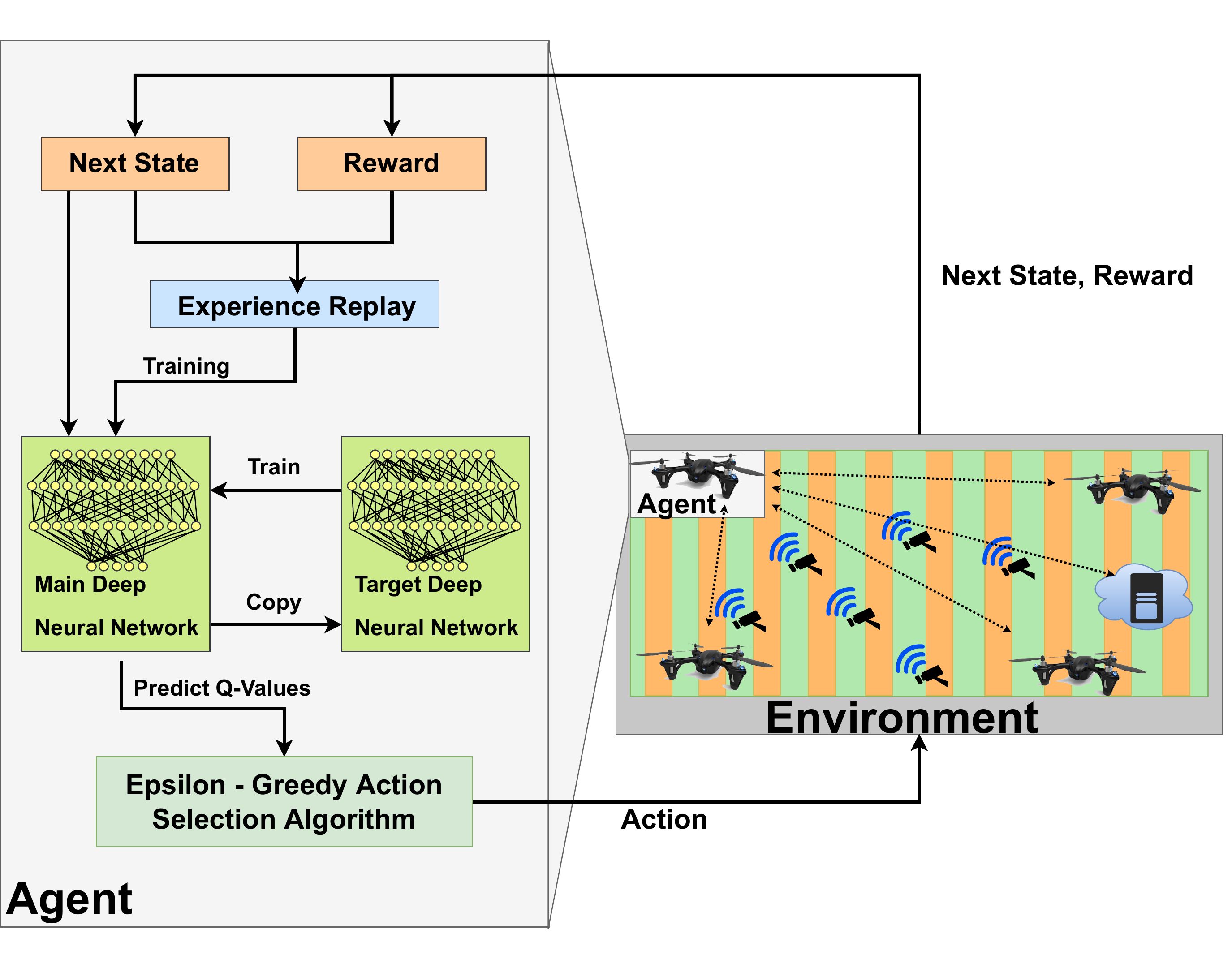}
    \caption{\label{fig:my_label} Deep Q-Learning Architecture}
    
\end{figure}
In the traditional Q-Learning, the Q-values are stored in a Q-table. When the agent needs to make a decision, it looks up the current state in the Q-table and selects the action with the highest Q-value. The Q-value measures the future cumulative discounted reward of the action at a given state. After the agent performs the selected action, the Q-value for that state-action pair is updated in the Q-table, and the agent moves on to another state. Due to the computer's finite amount of memory, Q-Learning's state space and action space are limited. 

With DQL, instead of looking up the Q-value in a Q-table, a DNN is used to predict the Q-values for each action at a given state. After the agent selects and performs the action, the agent's experience is collected. The experience is a tuple consisting of the agent's current state, next state, action, and reward. The experience is stored in a buffer called the experience replay, and the buffer is used to train the DNN. With more experience, the DNN becomes more accurate in predicting the Q-value of each action.   

A Markov decision process (MDP) framework provides a mathematical representation of a decision-making problem. In an MDP, there is an agent interacting with an environment through actions, and each action has an effect on the environment. After the agent performs the action in the environment, the environment returns a new state and reward to the agent, the agent must select a new action, and the process is repeated. The future state depends only on the current state and action. Finally, the agent needs to find a series of actions that will maximize the cumulative reward. 

Every UAV in the network will have its own MDP framework. In this problem, the UAVs are the agents and they receive tasks from the IoT devices, and they must decide where the tasks will be processed. After the UAVs send the tasks to the appropriate processing unit, the UAVs' battery levels change, the processing unit delays change, and these changes are reported back to the UAVs. The UAVs must select processing units that will minimize the number of deadline violations and energy consumption, which will in turn lead to the highest reward. The MDP is defined as follows:
\begin{itemize}
    \item \textbf{State:} The state consists of the offloaded task's type $k$, all processing unit delays $\Delta_{j'\in\mathcal{J^{+}}}$, the battery levels of every UAV $\Upsilon^{L}_{j'\in\mathcal{J}}$, and the transmission delays between every UAV and MEC device $\Delta^{j_1 \in\mathcal{J^+}t \in\mathcal{T}}_{j_2 \in\mathcal{J^+}}$. The state is defined as,
    \begin{flalign}
    \label{eq:DQLstate}
    \mathbb{S}= \{k, \Delta_{j'\in\mathcal{J^{+}}},  \Upsilon^{L}_{j'\in\mathcal{J}}, \Delta^{j_1 \in\mathcal{J^+} t \in\mathcal{T}}_{j_2 \in\mathcal{J^+}}\}. 
    \end{flalign}
    
    \item \textbf{Action:} The action is to choose a processing unit among the set of available processing units $j' \in \mathcal{J+}$, where the task will be computed. The tasks can be done by the local processing unit, the processing unit of a neighbouring UAV, or the MEC servers. Therefore the action is given as,
    \begin{flalign}
    \label{eq:action}
    \mathbb{A} = \{ x_{j' \in J'} \}.
    \end{flalign}
    
    \item \textbf{Reward:} The reward function is defined as the sum of two terms, the battery level reward ($\Upsilon^{L}_{j_{a}} - 1$), and the deadline violation reward  $(1-\mathbb{E}(v_{j_{a}})) + \mathcal{V}^{L}_{j_{a}} * \mathbb{E}(v_{j_{a}}))$.

    $\Upsilon^{L}_{j_{a}}$ rewards the agent for choosing an action that does not lead to a significant increase in energy consumption. $e$ refers to the energy consumption change threshold. $\mathcal{V}^{L}_{j_{a}}$ penalizes the agent for selecting an action that resulted in a deadline violation. If the deadline violation could have been prevented by offloading the task to a different processing unit, then the penalty is severe. If the deadline violation is inevitable, then the penalty is milder because there does not exist a better location to compute the task.
    
    \begin{flalign}
    \label{eq:rew1}
    &\mathbb{R} = (\Upsilon^{L}_{j_{a}} - 1) + (1-\mathbb{E}(v_{j_{a}})) + \mathcal{V}^{L}_{j_{a}} * \mathbb{E}(v_{j_{a}}) \\
    \label{eq:rew2}
	& \Upsilon^{L}_{j_{a}} =
	\begin{cases} 
	        2, & \text{if } \mathbb{E}(\Upsilon^{R}_{j_{a}}) - \max_{j'\in\mathcal{J}}(\mathbb{E}(\Upsilon^{R}_{j'})) \ge -e \\
            0,& \text{if } \mathbb{E}(\Upsilon^{R}_{j_{a}}) - \max_{j'\in\mathcal{J}}(\mathbb{E}(\Upsilon^{R}_{j'})) \le -2e \\
            1, & \text{otherwise,}
	\end{cases}\\
	\label{eq:rew3}
	& \mathcal{V}^{L}_{j_{a}} =
	\begin{cases} 
        -40, & \text{if } \mathbb{E}(v_{j_{m}}) = 0 \\
        -20,& \text{if } \mathbb{E}(v_{j_{r}}) = 0 \\
        -10,& \text{if } \exists j'\in(\mathcal{J}/(j_{r}\cup j_{a}))(\mathbb{E}(v_{j'})) = 0 \\
        -1, & \text{otherwise.}
	\end{cases}
	\end{flalign}
The values in (\ref{eq:rew2}) and (\ref{eq:rew3}) are selected such that the agents are reinforced to choose optimal battery and deadline values in fewer learning cycles. 
	\item \textbf{Policy:} We use the well-known epsilon-greedy action selection algorithm.
	
\end{itemize} 

In the following subsections, we explain the baseline schemes.

\subsection{Baseline Methods}
In order to investigate the effectiveness of the proposed method, it is compared with the methods presented in \cite{Nguyen2021}.
\par \subsubsection{Round Robin (RR)}
Every device with a processing unit in the network $j'\in \mathcal{J}^{+}$, is assigned an order from 1 to $J^{+}$. The current UAV will cycle through the ordered list to determine where to offload its task. 

\par \subsubsection{Highest Energy First (HEF)}
The UAVs regularly update each other on their current battery levels. The current UAV will first find the device with the highest remaining battery level. If the difference between the current energy level and the highest energy level is more than 1\%, then offload the task to the UAV with the highest energy level, otherwise, compute the task locally. Because MEC devices have unlimited power, we have to constrain the number of times tasks can be sent to MEC. Each MEC device has a  $ 1 / J^{+} $ chance of being selected.

\par \subsubsection{Lowest Queue Time and Highest Energy First (QHEF)} The UAVs regularly update each other on their current battery levels and queue times. First, the algorithm finds the minimum queuing time. Then the UAV finds the device that has the highest energy level and a queue time that is lower or equal to the minimum queue time. If the highest energy level is higher than the current energy level by a threshold, then the current UAV will offload the task to that device. Otherwise, the UAV will compute that task locally.

\subsubsection{Q-Learning} 
We used the Q-Learning algorithm presented in \cite{Nguyen2021}. The Q-Learning algorithm uses the action set defined in (\ref{eq:action}), reward function defined in (\ref{eq:rew1}), and epsilon-greedy policy. The Q-Learning algorithm's state is the same as (\ref{eq:DQLstate}), but without the transmission delays $\Delta^{j_1 \in\mathcal{J+} t \in\mathcal{T}}_{j_2 \in\mathcal{J+}}$.  

\section{Performance Evaluation}


\begin{table}
    \centering
    \caption{\label{tab:DQLParams} DQL  Parameters}
    \begin{tabular}{l|l}
    \textbf{Parameter} & \textbf{Value}\\
    \hline
    Number of Agents &  4\\
    \hline
    Batch Size & 500 \\
    \hline
    Experience Replay Size & 100000 \\
    \hline
    DNN Architecture & 
        \begin{tabular}{@{}p{\dimexpr 0.1\linewidth-1\tabcolsep}|
        p{\dimexpr 0.2\linewidth-2\tabcolsep}|
        p{\dimexpr 0.15\linewidth-2\tabcolsep}}
             Layer Type & Num. of Neurons & Activation Func.  \end{tabular} \\
    \hline
     & 
        \begin{tabular}{@{}p{\dimexpr 0.1\linewidth-1\tabcolsep}|
        p{\dimexpr 0.2\linewidth-2\tabcolsep}|
        p{\dimexpr 0.15\linewidth-2\tabcolsep}}
             Input & \hspace{4 mm} 10 & \hspace{2 mm} N/A
        \end{tabular} \\ 
     & 
        \begin{tabular}{@{}p{\dimexpr 0.1\linewidth-1\tabcolsep}|
        p{\dimexpr 0.2\linewidth-2\tabcolsep}|
        p{\dimexpr 0.15\linewidth-2\tabcolsep}}
             Hidden  & \hspace{4 mm} 32 & \hspace{1 mm} ReLU 
        \end{tabular} \\
     & 
        \begin{tabular}{@{}p{\dimexpr 0.1\linewidth-1\tabcolsep}|
        p{\dimexpr 0.2\linewidth-2\tabcolsep}|
        p{\dimexpr 0.15\linewidth-2\tabcolsep}}
             Hidden & \hspace{4 mm} 32 & \hspace{1 mm} ReLU
        \end{tabular} \\
    & 
        \begin{tabular}{@{}p{\dimexpr 0.1\linewidth-1\tabcolsep}|
        p{\dimexpr 0.2\linewidth-2\tabcolsep}|
        p{\dimexpr 0.15\linewidth-2\tabcolsep}}
             Output & \hspace{5 mm} 5 & \hspace{2 mm} N/A  
        \end{tabular} \\ 
             
    \hline
    DNN Loss Function & Mean Squared Error \\
    \hline
    DNN Optimization Function & Adam w/ Learning Rate of 0.001\\
    \hline
    \end{tabular}

\end{table}

\begin{table}
    \centering
    \caption{\label{tab:TaskParams} Task parameters from \cite{Nguyen2021}.}
    \begin{minipage}{0.45\textwidth}
    
    \begin{tabular}{l|@{}c@{}c@{}c@{}c@{}}
    Task Type &
    \begin{tabular}{p{0.7cm}|p{0.5cm}|p{1.3cm}|p{1.3cm}} $(1/\lambda)\footnote{Mean interarrival rate.}$ & $\alpha^{D}_{jt}$ & $\alpha^{P}_{jt} (UAV)$ & $\alpha^{P}_{jt} (MEC)$\\ \end{tabular} \\
    \hline
    Fire Detection & \begin{tabular}{p{0.7cm}|p{0.5cm}|p{1.3cm}|p{1.3cm}} 0.25s & 0.3s & 0.1s & 0.05s  \end{tabular} \\
    Pest Detection & \begin{tabular}{p{0.7cm}|p{0.5cm}|p{1.3cm}|p{1.3cm}} 0.25s & 0.8s & 0.5s & 0.25s \end{tabular} \\
    Growth Monitoring & \begin{tabular}{p{0.7cm}|p{0.5cm}|p{1.3cm}|p{1.3cm}} 0.5s & 5s & 0.1s & 0.05s \end{tabular}\\
    \hline
    \end{tabular}
    \end{minipage}
\end{table}

We used Simu5G, a 5G network simulator that runs on top of Omnet++ \cite{Nardini2020}, to simulate our smart farm network. In our simulation, we have four UAVs ($J = 4$), and one MEC device ($L = 1$). There are three task types: fire detection, pest detection, and growth monitoring. The task interarrival time is modeled as an exponential distribution. Each task type has a unique mean interarrival rate and processing time, and their values are presented in Table \ref{tab:TaskParams}.

The remaining battery level and delay violation results are the averages of ten runs with different seed values. For Q-Learning and Deep Q-Learning, a learning rate of 0.05 is assumed, and a discount value of 0.85 is considered. 

In order to compare our work with \cite{Nguyen2021}, we used their energy consumption model and parameters. We also made the same assumptions when it came to battery type, and hovering power consumption formula. We used (\ref{eq:encalc}) to model a UAV's energy level throughout our simulations. The values (in Watt-Hour) for each energy consumption parameter \footnote{We set the energy consumption level of idle and busy CPU periods to be at the level they would be if they ran for ten hours. This was done in order to showcase the performance of methods in terms of energy consumption in a limited amount of simulation time.} are as follows, the maximum battery capacity ($\Upsilon^{B}_{j'}$) is equal to 570, hovering ($\Upsilon^{H}_{j'}$) is equal to 211, the antenna is equal to 17, an idle processing unit is equal to 4320, and an active processing unit is equal to 12960 \cite{Nguyen2021}.

\par \subsection{Simulation Results} 

\par \subsubsection{Convergence}
\begin{figure}
    \centering
    \includegraphics[width=0.5\textwidth, height=6cm]{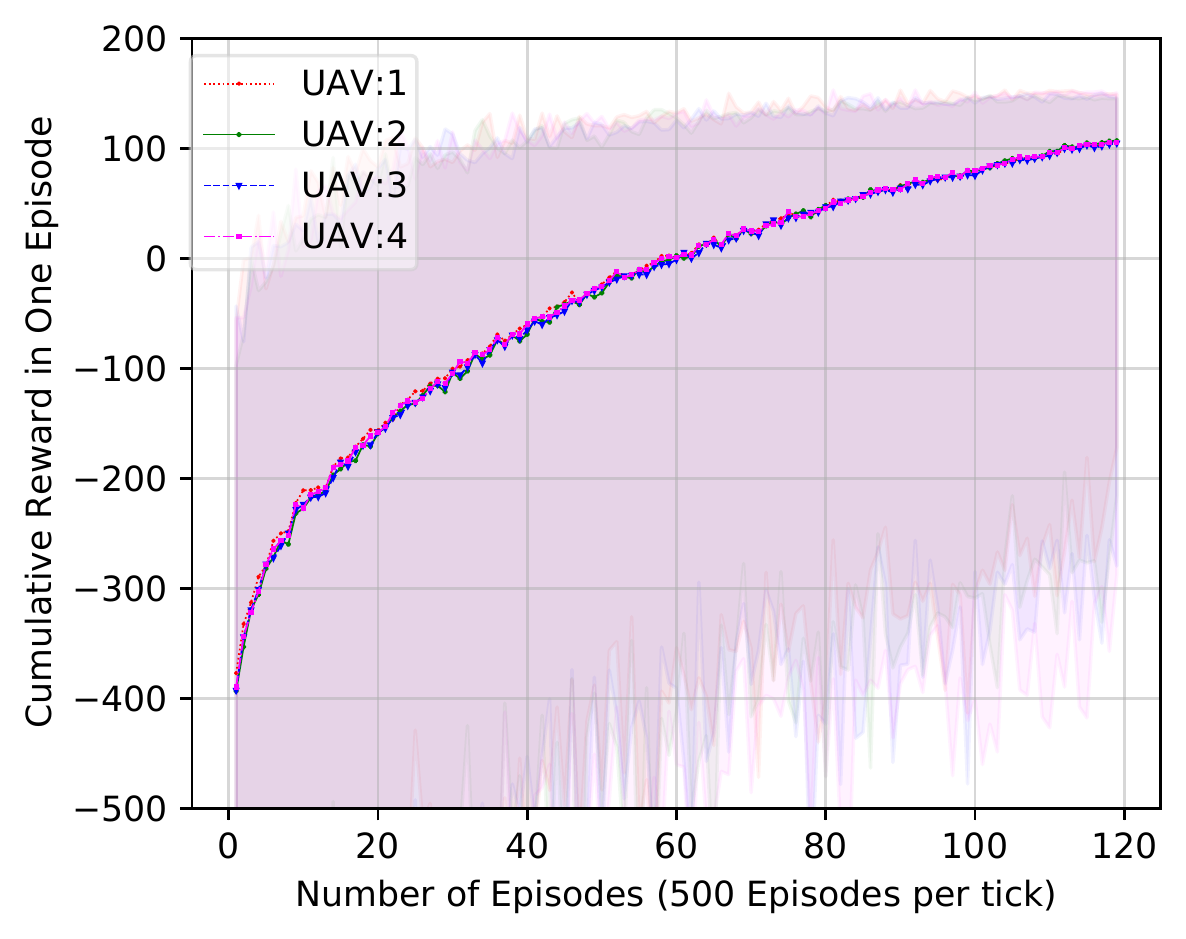}
    \caption{\label{fig:QL_Conv} Convergence of multi-UAV Q-Learning method}
\end{figure}
Fig. \ref{fig:QL_Conv} shows Q-Learning's cumulative reward for one episode, for 60k episodes. The rewards began converging to an average reward of 100 per episode after approximately 55k episodes. The solid lines represent the average cumulative reward for 500 or 100 episodes. The shaded area shows the variation of the average cumulative rewards. 
\begin{figure}
    \centering
    \includegraphics[width=0.5\textwidth, height=6cm]{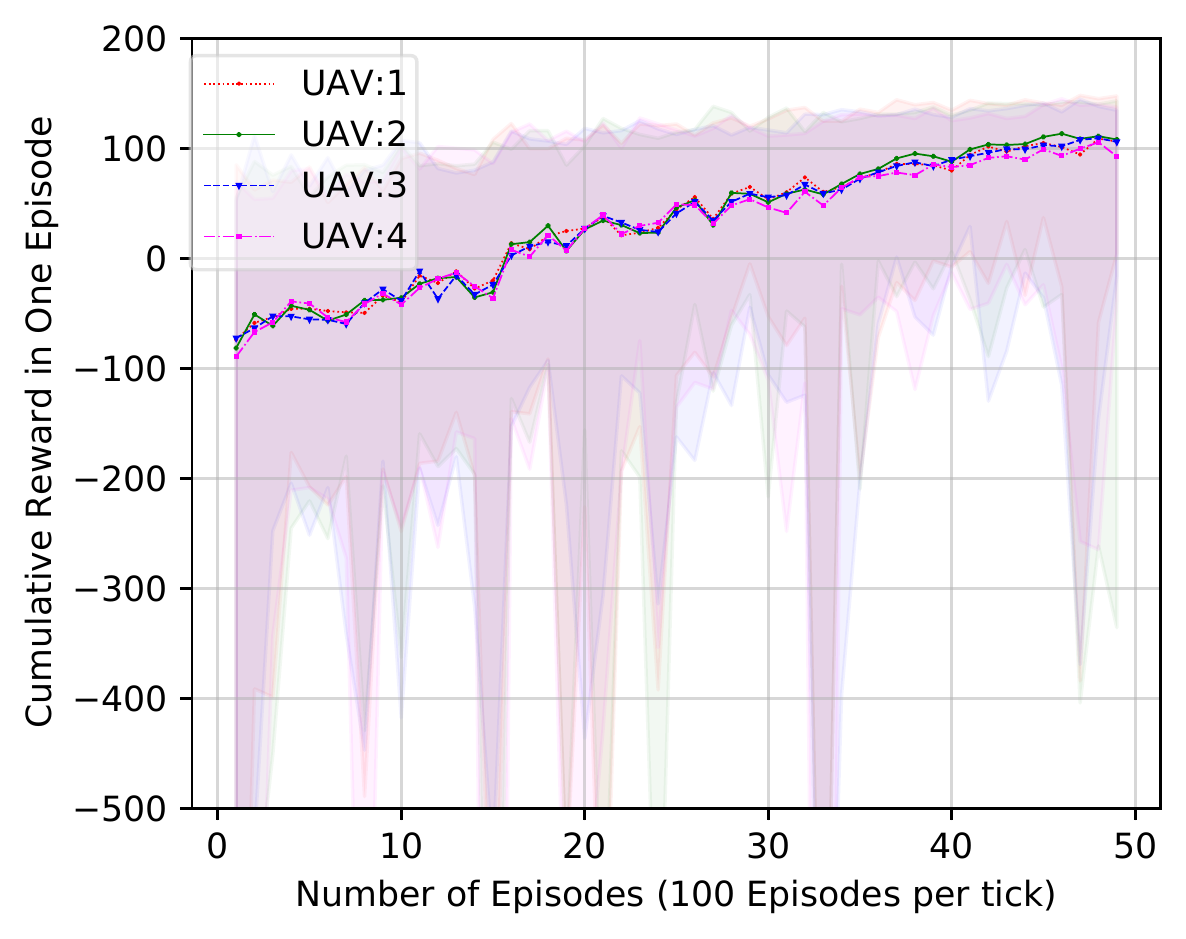}
    \caption{\label{fig:DQL_Conv} Convergence of multi-UAV DQL method}
\end{figure}
Fig. \ref{fig:DQL_Conv} shows DQL's cumulative reward for one episode, for 5k episodes. The rewards began converging to an average reward of 100 per episode after approximately 4200 episodes. 
Figures \ref{fig:QL_Conv} and \ref{fig:DQL_Conv} demonstrate that the Q-Learning algorithm requires approximately 13 times more episodes to reach convergence in comparison with the proposed DQL algorithm. The variation in cumulative reward in an episode in DQL decreased faster than the variation seen in Q-Learning. This is due to DQL's DNN constantly improving its ability to predict the Q-values. The variance is also caused by the varying total number of tasks in an episode. Each episode had a different random seed which influenced the starting point for task offloading for each IoT device, which in turn affects the total number of tasks in circulation.

\par \subsubsection{Remaining Battery Level}
\begin{figure}
    \centering
    \includegraphics[width=0.5\textwidth]{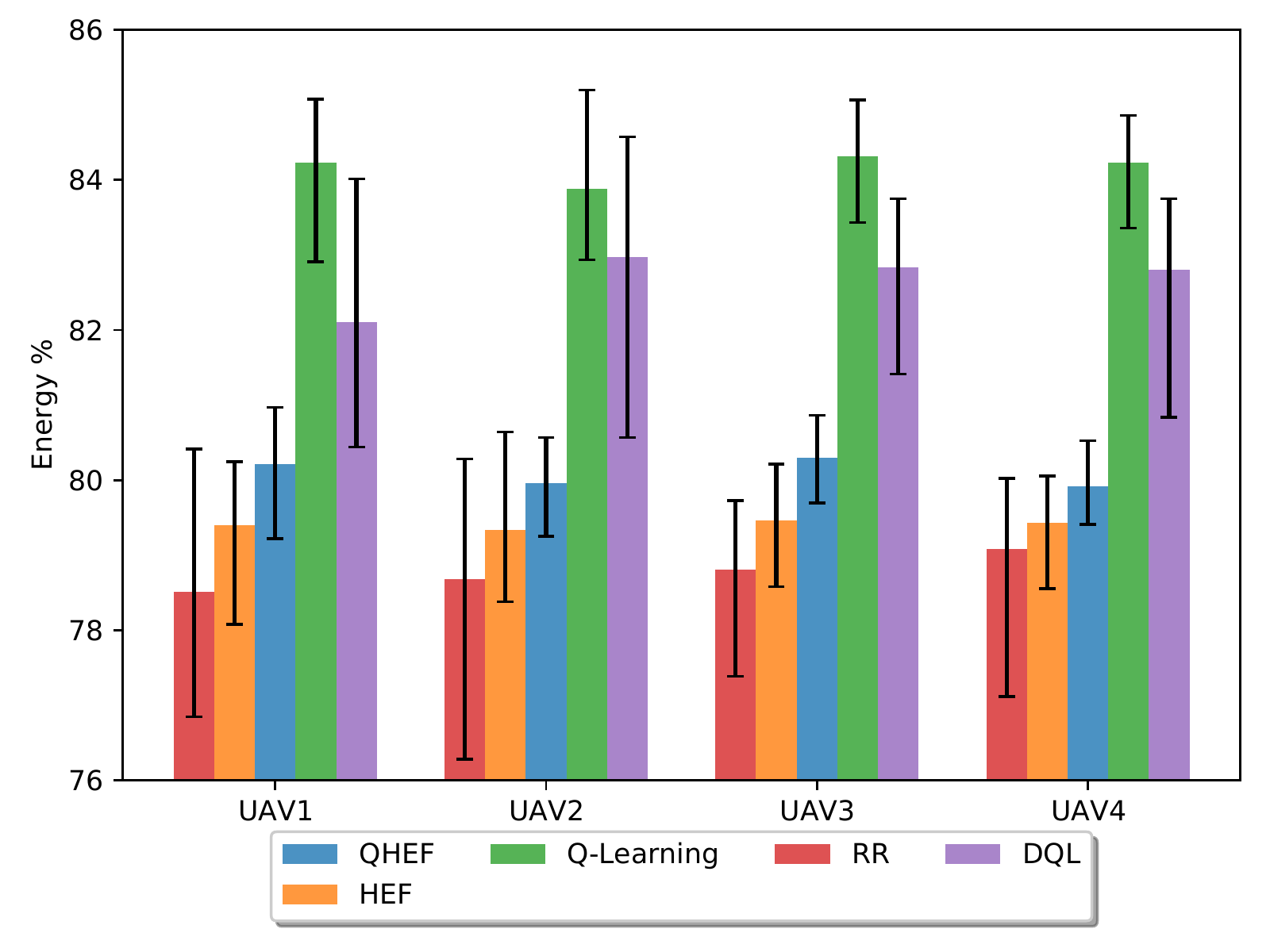}
    \caption{\label{fig:energy} Comparing the UAVs' remaining energy levels}
\end{figure}

The remaining battery level is defined as the percentage of energy that remains in the UAV's battery at the end of the simulation. This percentage indicates how long the UAV can remain hovering. A higher percentage corresponds to a long-lasting hover time. Fig. \ref{fig:energy} indicates that Q-Learning has the highest minimum remaining energy percentage. DQL is not far behind, it is approximately 2\% lower than  Q-Learning's minimum remaining energy percentage. The RR method had the lowest remaining energy percentage because it does not consider energy in its decision-making process.
Note that, both machine learning based techniques are compared after they reach convergence. Hence, DQL provides a close performance under a shorter convergence time.  
\par \subsubsection{Deadline Violations}
\begin{figure}
    \centering
    \includegraphics[width=0.5\textwidth]{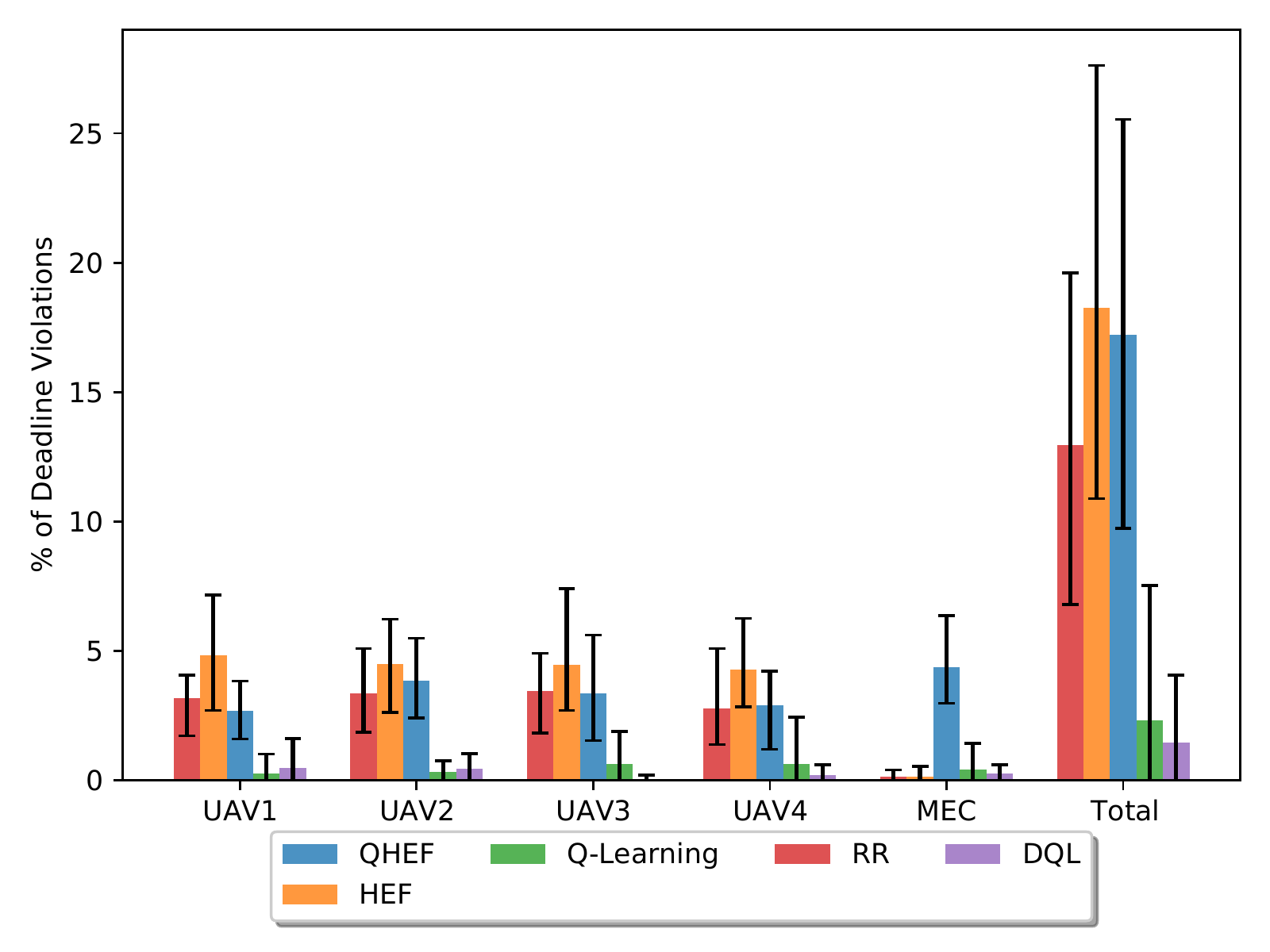}
    \caption{\label{fig:DV} Comparing the UAVs' deadline violation distribution}
\end{figure}

A deadline violation occurs when the processing unit has not completed the task before its deadline. (\ref{eq:DV}) can be used to determine if a deadline violation has occurred. Fig. \ref{fig:DV} illustrates the percentage of deadline violations that occurred at each node out of the total number of tasks that were generated. In terms of deadline violations, DQL is the best performing algorithm because it has the lowest total percentage of deadline violations. This is because DQL was able to consider all of the transmission delays between the UAVs and MEC device in its decision-making process. Nevertheless, Q-Learning has comparable performance, it has approximately 0.9\% more deadline violations than DQL. The HEF algorithm has the worst performance because it does not consider any type of delay or deadline in its decision-making process.The QHEF method also has poor performance because it offloaded too many tasks to the MEC device and increased the MEC device's queue time.  

\section{Conclusion and Future Works}
In this study, we presented a task distribution algorithm for a MEC assisted IoT-UAV smart farm network. We proposed a DQL-based algorithm to improve the convergence speed of an existing Q-Learning algorithm. The deep learning part of the algorithm also allowed us to include more observations into the state, therefore our decision-making algorithm had more information than Q-Learning. We investigated the proposed algorithm against four baseline algorithms RR, HEF, QHEF, and Q-Learning. The results demonstrated that the DQL algorithm is able to converge 13 times faster than Q-Learning. Finally, DQL had comparable results to Q-Learning when it came to remaining energy percentage and percentage of deadline violations. Therefore it is a more optimal solution for our joint optimization problem, with the ability to reach the optimal solution faster than Q-Learning. In the future, we plan to work on reducing the convergence further and also addressing scalability issues.

\section*{Acknowledgement}
This  work  is  supported  by  MITACS Canada Accelerate program under collaboration with Nokia Bell Labs.
\bibliography{tieraml}

\end{document}